\documentclass[twocolumn,astrosymb]{aastex701}

\usepackage{amsmath}

\begin{document}

\title{The Persistent Missing Mass Problem in Planet Formation}

\author[orcid=0000-0002-1228-9820,gname=Eve,sname=Lee]{Eve J.~Lee}
\affiliation{Department of Astronomy \& Astrophysics, University of California, San Diego, La Jolla, CA 92093-0424, USA}
\affiliation{Department of Physics, McGill University, 3600 rue University, Montr\'eal, QC H3A 2T8, Canada}
\email[show]{evelee@ucsd.edu}

\author[orcid=0000-0003-1827-9399]{William DeRocco}
\affiliation{Maryland Center for Fundamental Physics, University of Maryland, College Park, 4296 Stadium Drive, College Park, MD 20742, USA}
\email[]{}

\author[orcid=0000-0002-1032-0783]{Sam Hadden}
\affiliation{Canadian Institute for Theoretical Astrophysics, 60 St George St Toronto, ON M5S 3H8, Canada}
\email[]{}

\author[orcid=0000-0003-0395-9869]{B. Scott Gaudi}
\affiliation{Department of Astronomy, The Ohio State University, 140 West 18th Avenue, Columbus, OH 43210, USA}
\email[]{gaudi.1@osu.edu}

\begin{abstract}
Recent ground-based microlensing surveys suggest that our Galaxy may abound with small free floating planets, potentially up to $\sim$21 such planets per star. We explore the implication of such possibility on the mass budget for planet formation. When the microlensing planets, both bound and free-floating, are taken into account, along with the short-period planets, T Tauri disks have insufficient mass to source the mass of known planets, even if all the solids convert into planetary bodies. Younger Class 0/I disks can help resolve the problem but generally fall short of the required mass when variable planet formation efficiency from pebble or planetesimal accretion is taken into consideration. If the free-floating planet mass function is as bottom-heavy as reported, heavier Class 0/I disks may be necessary. Alternatively, free-floaters may preferentially form in the most massive disks around massive stars consuming the majority of the mass budget, leading to a decrease in the bound planet occurrence rate for higher mass stars, which is observed. Precise constraints on the bottom of planet mass function are necessary: a peaked mass function may eliminate the missing mass problem; by contrast, verifying a bottom-heavy function could spell a crisis in planet formation.

\end{abstract}

\section{Introduction} 
\label{sec:intro}

With more than 6000 detected exoplanets, our understanding of how planets form and evolve has advanced dramatically in the past decade, particularly for planets on short orbital periods. By comparison, beyond $\gtrsim$300 days, the population of small planets (super-Earths and smaller) remains unknown as they cannot be detected with present-day transit, radial velocity, or imaging techniques. At these intermediate periods, microlensing is an ideal planet detection method, uniquely capable of probing planets there, potentially down to terrestrial mass regimes \citep[e.g.,][]{Mao1991,Gould1991,Bennett1996,Bennett2002,Penny19,Zhu21}.

Ground-based microlensing surveys for bound planets generally indicate a bottom-heavy mass function for planet-to-star mass ratios of $q\gtrsim 10^{-4}$. For example, Microlensing Observations in Astrophysics (MOA)-II report a power-law slope of $dN/d\log{q} \propto q^{-1}$ (approximate) for mass ratios $q\gtrsim 10^{-4}$ \citep{Suzuki16}, whereas the Korean Microlensing Telescope Network (KMTNet) report a slope of $\sim -0.5$ for a power-law fit for mass ratios $q\gtrsim 10^{-4}$ \citep{Zang25}. For smaller mass ratios, MOA reports a break (a likely peak) at $q\sim 2\times10^{-4}$, whereas KMTNet prefers a double-Gaussian model for $dN/d\log{q}$ over a single power-law, with peaks at $q\sim 10^{-4.7}$ and $q\sim 10^{-2.6}$, and a dip at $q\sim 10^{-3.2}$ \citep[see][their Figure 3]{Zang25}. Regardless of the model used to fit the mass ratio distribution, both teams favor a plateau or a decrease in the abundance of small bound planets with mass ratios $q\lesssim 10^{-4}$ to $10^{-5}$. However, the lack of sensitivity of either survey to mass ratios $\lesssim 10^{-5}$ precludes the ability to definitively determine the shape of the mass function at lower masses. 

By contrast, these same microlensing surveys find a bottom-heavy mass function for candidate free-floating planet (FFP),\footnote{These microlensing events should be regarded as due to {\it candidate} FFPs, since they could instead be bound planets on sufficiently wide orbits that the signature of the host star is undetectable \citep{An05}. For the purposes of this study, our conclusions are unchanged whether these planets are widely bound or truly free-floating. We will therefore simply refer to them as FFPs.} with no evidence for a break or plateau at low masses. Using the longer baseline 9-yr microlensing survey of MOA-II, \citet{Sumi23} report the best-fit free-floating planet (FFP) power-law mass function with a slope of approximately -1 in $dN/\log{M_p}$ and a relatively high normalization, implying that there may exist roughly 21 FFPs per star down to $\sim 0.33M_\oplus$ \citep[see also][for earlier results supporting a more bottom-heavy mass function]{Mroz17} Using a sample of microlensing events from the KMTNet survey from 2016-2019, \cite{Gould2022} find a mass function with a slope and normalization consistent with \citet{Sumi23}. In both cases, the constraints on the shape and normalization of the FFP are derived from a handful (6--12) of short timescale events and thus are very uncertain. Furthermore, the lowest-mass FFP events in these samples are likely due to planets of mass $M_p \ga 0.1~M_\oplus$, so there are no constraints on the FFP mass function for planets substantially less massive than Earth. Therefore, similar to bound planet analyses, this lack of sensitivity to low mass objects implies the FFP mass function may be peaked rather than being strictly bottom-heavy \citep[see][their Figure 6]{Sumi23}.

These small FFPs are likely formed in protoplanetary disks then subsequently ejected (or scattered to wide orbits) from their host systems. The reported large number of FFPs therefore raises the question of whether typical planetary systems would have enough mass to create all the planets, both bound and free-floating. It has been pointed out that more evolved T Tauri disks do not contain enough solids to create the planets that have been detected and confirmed \citep[e.g.,][]{Greaves10,Najita14,Manara18}. While \citet{Mulders21} established that the mass discrepancy can be resolved by accounting for observational biases in the mass estimation of observed planetary systems (focusing on radial velocity and transit data), they emphasize that the Class II disk masses still fall short should the planet formation efficiency be less than 100\% (i.e., not all the disk solid material is incorporated into planets).

The initial stages of planet formation likely proceed by pebble accretion \citep[e.g.,][]{Ormel10,Lambrechts12} whereby the aerodynamic drag on small dust grains with Stokes number $\leq$1 aid their accretion onto larger bodies. The same aerodynamic drag shuttles the grains towards the inner regions of disks, and it has been established that the amount of dust that passes by the accreting body by this radial drift is greater than that accreted onto the body by an order of magnitude or more \citep[e.g.,][]{Ormel18,Lin18,Chachan23}, suggesting the planet formation efficiency is $\sim$10\%, although the exact value depends on the properties of the disk. 

We revisit the problem of mass budget in planet formation. Our study updates the previous literature by 1) focusing on the results of microlensing surveys which are sensitive to intermediate to wide (or unbound) orbits and thereby more fully capturing the {\it total} mass of planets in planetary systems; 2) using the measured masses of younger Class 0/I disks instead of T Tauri disks so as to properly capture the total {\it initial} mass budget \citep[see, e.g.,][]{Greaves11,Najita14,Chachan22}; and 3) self-consistently accounting for variable planet formation efficiency.

This paper is organized as follows. We describe our calculation methods in Section \ref{sec:methods} and present the results in Section \ref{sec:results}. Summary and implications are discussed in Section \ref{sec:discussion}.

\section{Methods}
\label{sec:methods}

Mass conservation dictates that the total sum of the bound and free-floating planets (their solid component) should be sourced from the initial solid mass within protoplanetary disks. We outline how we calculate the total required {\it solid} mass to account for the mass of the planets inferred from microlensing supplemented with that from transit/radial velocity surveys and compare the number to the measurements of solid disk masses.

\subsection{Required Total Mass from Exoplanet Observations}
 
Based on the MOA-II (2006--2014) survey towards the Galactic bulge, \citet{Sumi23} report a bottom-heavy (i.e., more planets at lower masses) mass function of FFPs:
\begin{equation}
    \frac{dN}{d\log M_p} = 2.18^{+0.52}_{-1.40}\left(\frac{M_p}{8\,M_\oplus}\right)^{-0.96^{+0.47}_{-0.27}},
    \label{eq:dNdMp_ffp}
\end{equation}
where $N$ is the number of planets per star and $M_p$ is the mass of the planetary body. In writing the above equation, we adopt the conservative microlensing detection limit (their CR2), equivalent to equation 15 of \citet{Sumi23}.  

Analyzing events from the KMTNet survey from 2016-2019, \citet{Gould2022} find a slope of $\sim$-0.9 to -1.2, with a normalization of $0.39 \pm 0.20$ at $M_p=38~M_\oplus$.  Adopting the \cite{Sumi23} slope of $-0.96$, the equivalent normalization at 8$M_\oplus$ is $1.74 \pm 0.87$, where the quoted uncertainty does not account for the uncertainty in the slope. Thus, the mass functions inferred by \citet{Sumi23} and \citet{Gould2022} are consistent. We will adopt the results from \citet{Sumi23} for their more definitive estimates.

For bound planets, we adopt the mass ratio function of \citet{Suzuki16} from the MOA-II (2007--2012) survey:
\begin{align}
    \frac{dN}{d\log q} &= \frac{A}{m}(s_{\rm max}^m-s_{\rm min}^m) \nonumber \\
    &\times
    \begin{cases}
        \left(\frac{q}{q_{\rm br}}\right)^n, & q > q_{\rm br}\\
        \left(\frac{q}{q_{\rm br}}\right)^p, & q < q_{\rm br}
    \end{cases}
    \label{eq:dNdq_bd}
\end{align}
where $q \equiv M_p/M_\star$, $M_\star$ is the mass of the host star, $s$ is the projected star-planet separation measured in angular Einstein radius ($s_{\rm max} = 5$, $s_{\rm min}=0.3$), $A = 0.61^{+0.21}_{-0.16}$, $q_{\rm br} = 1.7\times 10^{-4}$, $n=-0.93 \pm 0.13$, $p = 0.6^{+0.5}_{-0.4}$, and $m=0.49^{+0.47}_{-0.49}$. In writing the above equation, we adopt \citet{Suzuki16}'s fitting that fixes $q_{\rm br}$ and integrate out the dependence of their reported $d^2N/d\log sd\log q$ on $s$.

Since the reported uncertainties in \citet{Suzuki16} and \citet{Sumi23} on the slopes and normalizations of the respective fits do not reflect the correlations between parameters, they do not fully capture the true shape of the bound and free-floating planet confidence intervals. 
We instead performed linear fits to the upper and lower limits of the confidence intervals obtained from the fit posteriors of the two papers (N.~Koshimoto \& D.~Suzuki, private communication) which better represent the correlated errors between the various fit parameters. In addition, for FFPs, we adopt the posteriors to the ``broken power law'' to encompass the full range of uncertainties in the shape of the FFP mass function reported by \citet{Sumi23} but emphasize that we adopt the single power-law fit for our fiducial (``median'') FFP mass function.
Our approximated 1-$\sigma$ limits of the mass functions are as follows:
\begin{equation}
    \left.\frac{dN}{d\log M_p}\right|_{\rm FFP,lo} =  1.65 \times
    \begin{cases}
        \left(\frac{M_p}{2.76 M_\oplus}\right)^{1.3}, & M_p \leq 2.76M_\oplus \\
        \left(\frac{M_p}{2.76 M_\oplus}\right)^{-1.66}, & M_p > 2.76M_\oplus
    \end{cases}
    \label{eq:dNdMp_ffplo}
\end{equation}
\begin{equation}
    \left.\frac{dN}{d\log M_p}\right|_{\rm FFP,up} =  10 \times
    \begin{cases}
        \left(\frac{M_p}{2.76 M_\oplus}\right)^{-1.21}, & M_p \leq 2.76M_\oplus \\
        \left(\frac{M_p}{2.76 M_\oplus}\right)^{-0.8}, & M_p > 2.76M_\oplus
    \end{cases}
    \label{eq:dNdMp_ffpup}
\end{equation}
\begin{equation}
    \left.\frac{dN}{d\log M_p}\right|_{\rm BND,lo} =  0.5 \times
    \begin{cases}
        \left(\frac{M_p}{15 M_\oplus}\right)^{6.15}, & M_p \leq 15M_\oplus \\
        \left(\frac{M_p}{15 M_\oplus}\right)^{-0.8}, & M_p > 15M_\oplus
    \end{cases}
    \label{eq:dNdMp_bndlo}
\end{equation}
\begin{equation}
    \left.\frac{dN}{d\log M_p}\right|_{\rm BND,up} =  1.56 \times
    \begin{cases}
        \left(\frac{M_p}{15 M_\oplus}\right)^{0.61}, & M_p \leq 15M_\oplus \\
        \left(\frac{M_p}{15 M_\oplus}\right)^{-0.8}, & M_p > 15M_\oplus
    \end{cases}
    \label{eq:dNdMp_bndup}
\end{equation}
where `BND' stand for bound planets and `lo' and `up' correspond to lower and upper limits.

Using the results of the KMTnet survey, \citet{Zang25} find, unlike \citet{Suzuki16}, there is no clear evidence of a power-law break in the bound planet mass function. \citet{Zang25} attribute the source of the difference to the different planet identification methodology that allows them to be more sensitive to low mass planets than previous works \citep{Zang21}.
Most strikingly, \citet{Zang25} report a dip in the mass function at $q \sim 10^{-3}$, a potential signature of runaway gas accretion. It is not immediately obvious why the dip was not observed in other microlensing surveys. Such a dip in the mass function also remains in conflict with the radial velocity surveys that find the occurrence rate of Saturn-mass objects to be equivalent to that of Jupiter-mass objects \citep{Fulton21}. Nevertheless, we separately adopt the two mass ratio functions reported by \citet{Zang25}, considering both their single power-law:
\begin{equation}
    \left.\frac{dN}{d\log q}\right|_{\rm KMTnet} = 0.18\left(\frac{q}{10^{-4}}\right)^{-0.55},
    \label{eq:dNdq_zang_pl}
\end{equation}
and double Gaussian function:
\begin{align}
    &\left.\frac{dN}{d\log q}\right|_{\rm KMTnet} =  \nonumber \\
    &0.54\times 10^{-0.73(\log_{10}q+4.7)^2} + 0.058\times 10^{-1.8(\log_{10}q+2.6)^2}.
    \label{eq:dNdq_zang_dg}
\end{align}

The above mass functions encompass gas giants. To ensure an equivalent comparison with the solid mass budget available in protoplanetary disks, we convert the planet mass functions to the {\it solid} mass function. We choose to focus on solids because observations of protoplanetary disks provide more direct constraints on the solid content than on the gas content (dust continuum measurements are cheaper and more available than gas line measurements). Furthermore, the planetary mass-metallicity relation is much better constrained than disk solid-to-gas ratio. 

The empirical mass-metallicity relation in gas giants shows a large scatter \citep{Thorngren16,Chachan25}. While it has been proposed that the physical source of the relation and the scatter can be late-stage accretion of planetesimals \citep[e.g.,][]{Mousis09,Fortney13,Mordasini16}, the exact rate of solid accretion depends on the unknown dynamics of planetesimals \citep[e.g.,][]{Shibata23}. Another proposed hypothesis is merger by giant impact between multiple giants \citep{Ginzburg20}, with the scatter due to stochasticity in the number of impacts and the uncertain range of the critical core mass that triggers runaway accretion. Generally, the total solid mass of a given planet can be written as
\begin{equation}
    M_{\rm sld} = M_{\rm core} + fZ_\star M_{\rm env}
\end{equation}
where $M_{\rm core}$ is the mass of the core, $Z_\star$ is the stellar metallicity (assumed to be the local disk metallicity), $M_{\rm env}$ is the mass of the gaseous envelope, and $f$ is a free parameter that accounts for the uncertain physics that generates a scatter in planet envelope metallicity. \citet{Ginzburg20} fixed $f = 1$ and varied $M_{\rm core}$ whose mean scaling follows $M_{\rm core} \propto M_{\rm p}^{1/5}$ based on growth by gas accretion (which does not affect $M_{\rm core}$ but does affect the timescale to merger event) and the core mass doubling by mergers. 

To convert the planet mass function to the planet {\it solid} mass function, we adopt a simple methodology of fixing $M_{\rm core}$ and scaling down planets more massive than $M_{\rm core}$ to their corresponding $M_{\rm sld}$. Such an operation steepens the mass function beyond $M_{\rm sld} > M_{\rm core}$ compared to the total planet mass function. Following the updated empirical fit to the measured warm Jupiters' mass and their model-inferred total solid mass reported by \citet{Chachan25}, we adopt $M_{\rm core} = 14.72 M_\oplus$, $fZ_\star = 0.09\times U^{\log_{10}}(0.16, 3.33)$ where $U^{\log_{10}}(0.16, 3.33)$ is uniform distribution in base-10 logarithm between 0.16 and 3.33. We let $M_{\rm env}$ be the total mass of the planet in this high mass regime. Our choice of $fZ_\star$ is motivated by the visual inspection of the Figure 12 of \citet{Chachan25} (in particular the scatter in planet metallicity for planets more massive than 0.3 $M_{\rm Jup}$) and limits the scatter in planet metallicity only in the gaseous envelope, keeping $M_{\rm core}$ fixed.

\begin{figure}
    \centering
    \includegraphics[width=0.5\textwidth]{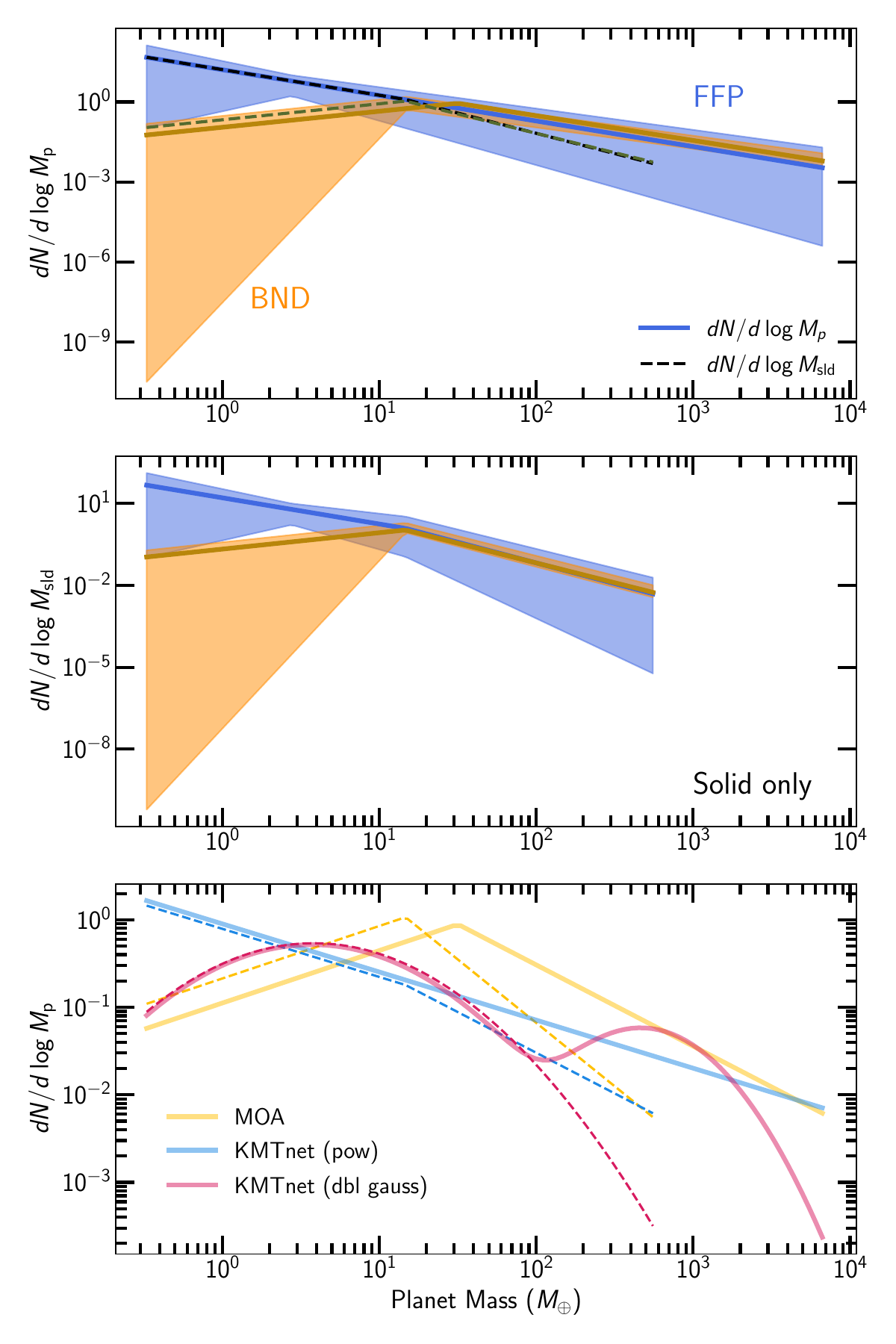}
    \caption{Planet mass functions reported in microlensing surveys. Top: results from the MOA survey for free-floating candidates \citep[blue;][]{Sumi23} and bound planets \citep[orange;][]{Suzuki16}. Solid lines represent the median while the shaded region correspond to the approximate 1-$\sigma$ error. Drawn in dashed are the mass functions of just the solid components of the planets. Middle: just the solid mass function for clarity. The maximum $M_{\rm sld}$ is scaled down from the maximum $M_p$ (see text).
    Bottom: comparison between the MOA \citep{Suzuki16} and KMTnet surveys \citep{Zang25} for the bound mass functions (solid) and their corresponding solid mass functions (dashed).}
    \label{fig:massfunction}
\end{figure}

\begin{deluxetable*}{lccccccccc}
\label{tab:sclfit}
\tabletypesize{\footnotesize}
\tablecaption{Solid mass function parameters}
\tablehead{\colhead{Type} & \colhead{a$_1$} & \colhead{a$_2$} & \colhead{a$_3$} & \colhead{$m_{\rm tr,1}/M_\oplus$} & \colhead{$m_{\rm tr,2}/M_\oplus$} & \colhead{A$_{m,1}$} & \colhead{A$_{m,2}$} & \colhead{N$_p$} & \colhead{Total M$_{\rm sld}/M_\oplus$}}
\startdata
FFP & -0.96 & -1.52 $\pm$ 0.18 & -- & 14.72 & -- & 1.22 & -- & 21.04 & 40.40 \\
FFP lo & 1.30 & -1.61 & -2.72 $\pm$ 0.30 & 2.76 & 14.72 & 1.64 & 0.11 & 0.94 & 3.34 \\
FFP up & -1.21 & -0.65 & -1.42 $\pm$ 0.07 & 2.76 & 14.72 & 10 & 3.33 & 48.83 & 98.95 \\
\hline
BND & 0.60 & -1.45 $\pm$ 0.05 & -- & 14.72 & -- & 1.07 & -- & 1.02 & 16.52 \\
BND lo & 6.15 & -1.50 $\pm$ 0.09 & -- & 14.72 & -- & 0.85 & -- & 0.30 & 9.84 \\
BND up & 0.61 & -1.45 $\pm$ 0.11 & -- & 14.72 & -- & 1.96 & -- & 1.84 & 30.19 \\
\hline
KMTnet (pow) & -0.55 & -1.01 $\pm$ 0.05 & -- & 14.72 & -- & 0.22 & -- & 1.30 & 7.46 \\
KMTnet (dbl) & \multicolumn{7}{c}{$0.58\times 10^{-[(\log_{10}(m/M_\oplus)-\log_{10}(3.61\pm 0.12))^2/2(0.80\pm 0.03)^2]}$} & 0.74 & 13.18\\
\enddata
\tablecomments{Column 1: FFP and BND each correspond to free-floating and bound planet mass functions, while `lo' and `up' refer to lower and upper limits, respectively. Columns 2--8: mass function parameters defined in equations \ref{eq:dNdmsld1} and \ref{eq:dNdmsld2}. Column 9: number of planets per star. Column 10: total solid mass in planets per star.}
\end{deluxetable*}

We draw a distribution of $10^5$ planets from each of the total planet mass function and apply the random generation of $M_{\rm sld}$ for planets more massive than $M_{\rm core}$. We repeat the step 1000 different instances to find the mean power-law scaling of the solid mass function beyond $M_{\rm core}$ using a least squares fit. Empirically, such an exercise creates a small pile-up of planets near $M_{\rm core}$ which is an artifact of our choice of expression for $M_{\rm sld}$ as it is formally discontinuous at $M_{\rm sld} = M_{\rm core}$. Given that the goal of this exercise is simply to get the power-law scaling at the high mass end, we enforce our final $dN/d\log M_{\rm sld}$ to be continuous at $M_{\rm core}$ and solve for the normalization factor(s) so that the {\it number} of planets per star remains the same over $M_p \in [0.33, 6660]M_\oplus$ (equivalently $M_{\rm sld} \in [0.33, 554.72]M_\oplus$):
\begin{equation}
    \frac{dN}{d\log m_{\rm sld}} = A_{m,1} \times
    \begin{cases}
        \left(\frac{m_{\rm sld}}{m_{\rm tr,1}}\right)^{a_1}, & m_{\rm sld} \leq m_{\rm tr,1}\\
        \left(\frac{m_{\rm sld}}{m_{\rm tr,1}}\right)^{a_2}, & m_{\rm sld} > m_{\rm tr,1}
    \end{cases}
    \label{eq:dNdmsld1}
\end{equation}
in case of a single break (e.g., when $dN/dlogM_p$ is a single power law or the location of its power law break occurs at a mass greater than $M_{\rm core}$) and
\begin{equation}
    \frac{dN}{d\log m_{\rm sld}} = 
    \begin{cases}
        A_{m,1}\left(\frac{m_{\rm sld}}{m_{\rm tr,1}}\right)^{a_1}, & m_{\rm sld} \leq m_{\rm tr,1}\\
        A_{m,1}\left(\frac{m_{\rm sld}}{m_{\rm tr,1}}\right)^{a_2}, & m_{\rm tr,1} < m_{\rm sld} \leq m_{\rm tr,2}\\
        A_{m,2}\left(\frac{m_{\rm sld}}{m_{\rm tr,2}}\right)^{a_3}, & m_{\rm sld} > m_{\rm tr,2}
    \end{cases}
    \label{eq:dNdmsld2}
\end{equation}
in case of a double break (e.g., when a power law break in $dN/dlog M_p$ occurs at a mass less than $M_{\rm core}$). In writing the above expressions, we set $m \equiv M_{\rm sld}/M_\oplus$. To convert the planet-star mass ratio into planet mass in the bound planet mass function, we use the average stellar mass 0.56$M_\odot$ derived from the stellar initial mass function employed by \citet{Sumi23}:
\begin{equation}
    \frac{dN}{d\log m_\star} =
    \begin{cases}
        A_1 m_\star^{-1.32}, & 0.86 < m_\star \leq 120\\
        A_2 m_\star^{-0.13}, & 0.08 < m_\star \leq 0.86\\
        A_3 m_\star^{0.58}, & 3\times 10^{-4} < m_\star \leq 0.08
    \end{cases}
    \label{eq:IMF}
\end{equation}
where $m_\star \equiv M_\star/M_\odot$. Setting $A_3 = 1$ and ensuring the IMF is continuous, $A_1 = 0.139$ and $A_2 = 0.166$. 
In case of the double Gaussian function of KMTnet survey, we enforce a single Gaussian fit to the resulting $dN/d\log m_{\rm sld}$. The parameters corresponding to each $dN/dl\log m_{\rm sld}$ are summarized in Table \ref{tab:sclfit}. The total number of planets per star $N_p$ is obtained by integrating $dN/d\log M_p$ over $\log M_p$ from $M_p=0.33M_\oplus$ to 6660$M_\oplus$ and the total solid mass of planets per star is obtained by integrating $m_{\rm sld} dN/d\log m_{\rm sld}$ over $\log m_{\rm sld}$ from $m_{\rm sld}=$ 0.33 to 554.72. Figure \ref{fig:massfunction} illustrates the mass functions we use, both $dN/d\log M_p$ and $dN/d\log M_{\rm sld}$, showing the steeper slope of the solid mass function compared to total mass function at the high mass end.

Mass functions reported from microlensing surveys are convolved over the full stellar initial mass function, so the dependence of the planet occurrence rate on stellar type is in a sense corrected for. On the other hand, the microlensing technique is most sensitive to planets at a few AU and beyond and not to planets at close-in distances. To account for the mass of planets inside $\sim$1 AU, we look to radial velocity and transit data. \citet{Dai20} report the minimum mass extrasolar nebula constructed from the California-Kepler Survey (CKS) with precise stellar characterization, focusing on small planets (1--4$R_\oplus$). While the baseline sample is from {\it Kepler} transit data, planet masses were taken from radial velocity or transit timing variation measurements where available. For targets without mass measurements, planet radii were converted using a number of mass-radius relationship reported in the literature, accounting for the systematic differences between them (see section 3.1 of \citealt{Dai20} for more detail). Their final reported mass distribution (corrected for detection biases) is
\begin{equation}
    \Sigma_{\rm inner} = 50^{+32}_{-19}\,{\rm g\,cm^{-2}}\left(\frac{a}{1\,{\rm au}}\right)^{-1.76\pm 0.07} \left(\frac{M_\star}{M_\odot}\right)^{1.39\pm 0.29},
\end{equation}
where $a$ is orbital distance.

The CKS sample used in \citet{Dai20} spans $a \in [0.02, 1]$ au and $m_\star \in [0.6, 1.4]$. The total planet mass inside 1 au per star for stars in their adopted mass range is then
\begin{equation}
    M_{\rm p, inner} = \frac{2\pi\int^{1.4}_{0.6}\int^{1}_{0.02} \Sigma_{\rm inner} a_{\rm 1 au} da_{\rm 1 au} \frac{dN}{d\log m_\star} \frac{d m_\star}{m_\star}}{\int^{1.4}_{0.6} \frac{dN}{d\log m_\star} \frac{d m_\star}{m_\star}}
\end{equation}
which evaluates to 26.3$M_\oplus$ where $a_{\rm 1 au} \equiv a/1\,{\rm au}$. 

The lower limit on the host stellar mass in the CKS sample is close to but greater than the average stellar mass of the microlensing survey. The occurrence rate of small planets around cooler M dwarfs is reported to be at least 3 times larger than that around FGK dwarfs \citep[e.g.,][]{Dressing15,Mulders15}. Correcting for stellar binarity can account for half of this enhancement \citep{Moe21} and further correcting for stellar-mass dependent planet multiplicity \citep{Yang20} is expected to bring the M dwarf planet occurrence rate into better agreement with that of FGK dwarfs \citep{Zhu21}. Data from Transiting Exoplanet Survey Satellite (TESS) show a hint of a decrease in the occurrence rate of small planets around stars lighter than $\sim$0.3$M_\odot$ \citep[e.g.,][]{Brady22,Ment23}. If we therefore assume that the stellar-mass dependence on $\Sigma_{\rm inner}$ extends down to stars of mass $M_\star = 0.3 M_\odot$, $M_{\rm p,inner}$ evaluates to 17.7$M_\oplus$.

\subsection{Comparison to Disk Masses}

Although it has been classically thought that planet formation begins in the T Tauri stage, the estimated disk masses at such evolved stage are too small to account for typical exoplanetary systems \citep[e.g.,][]{Greaves10,Najita14,Manara18}, suggesting the assembly of planet-building blocks may have started earlier during the Class 0/I stage \citep[e.g.,][]{Greaves11,Najita14}. \citet{Mulders21} find that the disk masses at T Tauri stage are comparable to their accounting of the total masses locked in planets from transit and RV surveys, once the latter are corrected for detection biases. However, their accounting did not consider bound planets detected by microlensing, which are generally distinct from those detected by RV and transit surveys. In addition, constraints on the frequency and mass function of free-floating planets from microlensing surveys were not available at the time of the \citet{Mulders21} study. 

Even considering only the planets detected by RV and transits, sourcing these planets from the measured solid masses of T Tauri disks would require the planet formation efficiency to be 100\%. Theoretical studies of core coagulation by pebble accretion---where planet formation efficiency $\epsilon_{\rm eff}$ is defined as the rate of pebble accretion divided by the mass flux of radial drift---report $\sim$10\% efficiency, with the exact value depending on the particle Stokes number, the local disk properties, and the planet mass \citep[e.g.,][]{Ormel18,Lin18,Chachan22}.

Before we provide a quantitative discussion of planet formation efficiency, we first describe the measurements of disk masses that we will adopt to compare with the total required dust mass to create the planets we observe. \citet{Manara23} present a comprehensive collection of T Tauri disks and their measurements collected from literature. Among these disks, only the stars (and brown dwarfs) in the mass range of 0.032 to 3.99 $M_\odot$ (the maximum stellar mass here corresponds to the maximum stellar mass in the entire sample) host disks with dust mass greater than 0.33 $M_\oplus$, our lower limit on $M_{\rm sld}$. To take account of the {\it initial} disk mass budget, we consider young (Class 0/I) disks. For these young disks, we use the collection of observations in the Orion region presented by \citet{Tobin20}. We adopt their Very Large Array (VLA) measurements (9 mm) of disk dust as the dust emission is less optically thick at longer wavelengths of VLA \citep{Zhu2019,Viscardi25}. The median disk mass is $\sim$200$M_\oplus$ (compared to the median mass of $\sim$3$M_\oplus$ in T Tauri disks) which can vary by order unity factors depending on different clusters \citep[e.g.,][]{Tychoniec20}. We revisit the matter of Class 0/I disk masses in Section \ref{ssec:timing}.

\subsubsection{Planet Formation Efficiency from Pebble Accretion}

Following the procedure of \citet{Chachan23}, we compute the required disk mass to amass a target planet whose total solid mass is $m_t$ as
\begin{equation}
    M_{\rm req} = \int^{m_t}_{m_0} \frac{1}{\epsilon_{\rm eff}}dm
    \label{eq:mreq_step}
\end{equation}
where $m_0$ is the mass of the embryo which we vary between $10^{-5} M_\oplus$ and $10^{-2} M_\oplus$.
Under pebble accretion, $\epsilon_{\rm eff}$ is a function of planet mass and the properties of the disk, which in turn depend on the stellar mass.
We fix the particle Stokes number St to $10^{-3}$ and turbulent $\alpha_t = 10^{-4}$ motivated by their inferred values from the analysis of ringed protoplanetary disks \citep[e.g.,][]{Lee24}. We further fix the orbital distance at 3 au where the sensitivity to microlensing planets is the highest; by doing so, we implicitly assume all free-floating planets originated from $\sim$3 au, which coincides with the distance beyond which planets more massive than Neptune can be ejected through scattering by equal mass neighbors \citep[e.g.,][]{Hadden25}.

We adopt the calculation of $\epsilon_{\rm eff}$ laid out in \citet{Chachan23}, summarizing only the key points here. First, the planet formation efficiency is defined as
\begin{equation}
    \epsilon_{\rm eff} = \frac{\dot{M}_{\rm acc}}{\dot{M}_{\rm drift}},
    \label{eq:eff_def}
\end{equation}
where $\dot{M}_{\rm acc}$ is the pebble accretion rate, $\dot{M}_{\rm drift} \equiv 2\pi a v_r \Sigma_{\rm sld}$ is the radial drift rate of the solids, $\Sigma_{\rm sld}$ is the solid surface density, and $v_r$ is the radial velocity of the solids defined as
\begin{equation}
    v_r = -\frac{3}{2} \frac{c_s^2}{v_k}\left[\alpha_{\rm t} + \frac{2}{3}|\gamma|{\rm St}\right],
    \label{eq:v_r}
\end{equation}
where $c_s \equiv \sqrt{k T / \mu m_H}$ is the sound speed, $k$ is the Boltzmann constant, $T$ is the disk midplane temperature, $\mu \equiv 2.37$ is the gas mean molecular weight, $m_H$ is the mass of the hydrogen atom, $v_k \equiv \sqrt{GM_\star/a}$ is the Keplerian orbital velocity, $G$ is the gravitational constant, $\gamma \equiv {\rm d \, ln}P_{\rm g} / {\rm d \, ln}a$, and $P_{\rm g}$ is the gas pressure. Gas pressure can be written as $P_{\rm g} = \Sigma_{\rm g} c_s \Omega_k$ where $\Omega_k \equiv \sqrt{GM_\star/a^3}$ is the Keplerian orbital frequency. We adopt $\Sigma_{\rm g} \propto a^{-1}$ motivated by the MAPS survey \citep{Zhang21}. For the disk temperature, we adopt the irradiation-heated disk 
\begin{equation}
    T = 150\,{\rm K} \left(\frac{a}{1\,{\rm au}}\right)^{-3/7} \left(\frac{M_\star}{M_\odot}\right)^{2/7},
    \label{eq:Tdisk}
\end{equation}
which derives from our adoption of $L_\star \propto M_\star^{1.5}$ where $L_\star$ is the stellar luminosity \citep{Dotter16,Choi16}, and the normalization is taken from the visual inspection of \citet{DAlessio98}, their Figure 2.
With our choice, $\gamma = -19/7$. Physically, the first term in equation \ref{eq:v_r} represents the coupling to the radial diffusive motion of the gas whereas the second term corresponds to the aerodynamic drag \citep{Weidenschilling77,Nakagawa86}. In writing equation \ref{eq:v_r}, we have taken the limit of St $\ll 1$ which is valid for our choice of St.

The pebble accretion rate $\dot{M}_{\rm acc}$ takes different forms depending on the accretion regime \citep[see][for a review]{Ormel17}. First, the accretion velocity (the velocity at which pebbles enter the accretion cross section) can be headwind (accretion velocity set by headwind drag) or shear (accretion velocity set by the relative orbital shear velocity between the pebble and the accreting body) dominated.\footnote{The velocity can alternatively be turbulence-dominated as well but for our chosen $\alpha_t$, particle turbulent velocity will always be sub-dominant \citep[see, e.g.,][]{Lin18,Rosenthal18}.} Second, the accretion can be three-dimensional (3D; the accretion cross section is smaller than the solid disk scale height) or two-dimensional (2D; the accretion cross section is larger than the solid disk scale height). 

Following \citet{Lin18} (see also \citet{Ormel10}), we find that for our adopted parameters, low-mass seeds always begin their accretion in the 3D regime.  Once the seeds accrete enough mass to reach the 2D accretion regime, accretion velocities switch to being shear-dominated. The mass at which the mode of accretion transitions from three- to two-dimensional can be derived as follows. We start with the condition for pebble accretion $v_{\rm acc} = 4GM_p/R_{\rm acc}^2 ({\rm St}/\Omega_k)$ where $R_{\rm acc}$ is the accretion radius and $v_{\rm acc}$ is the velocity at which solids enter $R_{\rm acc}$. Under the shear regime, $v_{\rm acc} = (3/2)R_{\rm acc}\Omega_k$. Plugging $v_{\rm acc}$ into the condition for pebble accretion, one can obtain $R_{\rm acc}$ and the transition mass is obtained following the condition for 2D accretion, $R_{\rm acc} > H_d \equiv (c_s/\Omega_k)\sqrt{\alpha_t/(\alpha_t + {\rm St})}$ where $H_d$ is the scale height of the solid disk. The transition mass is then
\begin{align}
    m_{\rm 2D, sh} &= \frac{3}{8} {\rm St}^{-1} \left(\frac{\alpha_t}{\alpha_t + {\rm St}}\right)^{3/2}\left(\frac{c_s}{v_k}\right)^3 \frac{M_\star}{M_\oplus} \nonumber \\
    &\equiv m_{\rm 2D, sh, 0} m_\star^{-1/14},
    \label{eq:m_2d_sh}
\end{align}
where $m_{\rm 2D,sh,0} = 126.27$ for our chosen parameters.
Below $m_{\rm 2D,sh}$, the accretion efficiency is
\begin{align}
    \epsilon_{\rm 3D} &= \dfrac{8}{3\pi}\left(\frac{M_p}{M_\star}\right)\left(\frac{v_k}{c_s}\right)^3{\rm St}\left(1+\frac{\rm St}{\alpha_t}\right)^{1/2}\bigg(\alpha_t+\dfrac{2}{3}|\gamma|{\rm St}\bigg)^{-1} \nonumber \\
    &\equiv \epsilon_{\rm 3D, 0} m m_\star^{1/14},
    \label{eq:eps_3d}
\end{align}
where we assume $M_p = M_{\rm sld}$. For a given stellar mass, the efficiency reaches 1 at
\begin{equation}
    m_{\rm 3D=1} = \epsilon_{\rm 3D,0}^{-1} m_\star^{-1/14} \sim 8.33 m_\star^{-1/14},
\end{equation}
which implies that at the minimum $m_\star=0.032$, $m_{\rm 3D=1} = 10.65$ and at the maximum $m_\star=3.99$, $m_{\rm 3D=1}=7.55$. For all stellar masses under our consideration, $m_{\rm 3D=1} < m_{\rm 2D,sh}$ so $\epsilon_{\rm eff} = \epsilon_{\rm 3D}$ for $m < m_{\rm 3D=1}$ and for higher masses, $\epsilon_{\rm eff}=1$ regardless of which regime the accretion is formally in since $\epsilon_{\rm eff}$ cannot physically exceed 1.

We can therefore compute the required dust mass {\it per star}, weighted by the planet and stellar mass functions as follows:
\begin{align}
    &\langle M_{\rm req} \rangle = N_\star^{-1} \times \nonumber \\
    &\int^{m^{\rm max}_{\star}}_{m^{\rm min}_{\rm \star}} \frac{dN}{d\log m_\star} d\log m_\star\left[\int^{m_{\rm 3D=1}}_{m_{\rm min}}\int^{m_t}_{m_0} \frac{1}{\epsilon_{\rm 3D}}dm \right. \nonumber \\
    &\left.+ \int^{m_{\rm max}}_{m_{3D=1}} \left(\int^{m_{t,3D=1}}_{m_0} \frac{1}{\epsilon_{\rm 3D}} + \int^{m_t}_{m_{t,3D=1}} \right)dm \right]\frac{dN}{d\log m_t}d\log m_t,
\end{align}
where $m_{\rm min} = 0.33$, $m_{\rm max} = 554.72$, and
\begin{equation}
    N_\star \equiv \int^{m^{\rm max}_{\star}}_{m^{\rm min}_{\star}} \frac{dN}{d\log m_\star} d\log m_\star \sim 0.32,
\end{equation}
for our adopted stellar initial mass function (equation \ref{eq:IMF}).
In words, we compute $M_{\rm req}$ for a given $m_t$ from equation \ref{eq:mreq_step} for all the planets per star over a solid planet mass function $dN/d\log m_t$ and {\it average} the stellar mass dependence of $\epsilon_{\rm ff}$ over the stellar mass function $dN/d\log m_\star$. The obtained $\langle M_{\rm req} \rangle$ are listed in Table \ref{tab:Mreq}.

\begin{deluxetable*}{lcccccccc}
\label{tab:Mreq}
\tabletypesize{\footnotesize}
\tablecaption{Required Mean Total Solid Mass Per Star}
\tablehead{\colhead{$m_0$} & \colhead{FFP} & \colhead{FFP lo} & \colhead{FFP hi} & \colhead{BND} & \colhead{BND lo} & \colhead{BND hi} & \colhead{KMTnet (pow)} & \colhead{KMTnet (gauss)}}
\startdata
Pebble accretion\\
$10^{-5}$ & 2234.85 & 109.18 & 5086.29 & 133.96 & 45.91 & 243.44 & 147.18 & 90.61 \\
$10^{-2}$ & 884.61 & 48.64 & 1999.19 & 68.82 & 26.36 & 125.21 & 63.64 & 42.62 \\
\hline
Planetesimal accretion\\
$10^{-4}$ & 106.63 & 7.98 & 282.89 & 64.33 & 41.24 & 117.64 & 27.79 & 19.35 \\
\hline
\enddata
\tablecomments{All masses are in $M_\oplus$.}
\end{deluxetable*}

\subsubsection{Other assembly processes}

In the previous section, we have limited our consideration to assembly by pebble accretion. Planets can gain their mass by alternative processes including planetesimal accretion and giant impacts. Quantifying the efficiency at which the planetesimals are incorporated into planets is challenging owing to the unknown random velocity of the planetesimals that can change the effective collisional cross section (i.e., the gravitational focusing factor). We adopt here a simple prescription of planetesimal accretion efficiency based on the consideration of gravitational scattering of planetesimals by the accreting planets.

Large-scale scattering of planetesimals is expected when the perturber (i.e., the growing planet) is massive enough for the Safronov number to be 1:
\begin{equation}
    \left(\frac{v_{\rm esc,p}}{2v_{\rm k}}\right)^2 = \frac{M_p}{M_\star}\frac{a}{R_p} \sim 0.2 \left(\frac{M_p}{M_\oplus}\right)^{3/4}\left(\frac{M_\odot}{M_\star}\right)\left(\frac{a}{3\,{\rm AU}}\right)
\end{equation}
where $v_{\rm esc,p}$ is the surface escape speed of a perturber, and $R_p$ is the radius of the perturber, where we use $R_p = R_\oplus (M_p/M_\oplus)^{1/4}$ \citep{Valencia06}. At 3 au, around 0.56$M_\odot$ star, the Safronov number reaches 1 for $M_p \sim 3.65 M_\oplus$. We may therefore expect the planet formation efficiency under planetesimal accretion to be less than unity for planets more massive than $\sim$3.65$M_\oplus$ (unlike pebble accretion where $\epsilon_{\rm eff} < 1$ for lighter planets). 

We write the mass beyond which the Safronov number exceeds 1 as $m_s$ and take a reduced planet formation efficiency by the scattering planetesimals $\epsilon_{\rm sc}=0.2$ \citep{Shibata23}. Around the maximum stellar mass $M_\star = 3.99M_\odot$, $m_s \sim 50$. However, our adoption of solid mass function implicitly assumes that planets of solid mass greater than 14.72$M_\oplus$ become gas-rich which will be massive enough to scatter the planetesimals so we set the maximum $m_s$ to be 14.72$M_\oplus$ for all stellar masses. 

We can then write the required dust mass per system under planetesimal accretion as
\begin{align}
    &\langle M_{\rm req} \rangle = N_\star^{-1} \times \nonumber \\
    &\int^{m^{\rm max}_{\star}}_{m^{\rm min}_{\rm \star}} \frac{dN}{d\log m_\star} d\log m_\star\left[\int^{m_s}_{m_{\rm min}}\int^{m_t}_{m_0} dm \right. \nonumber \\
    &\left.+ \int^{m_{\rm max}}_{m_s} \left(\int^{m_s}_{m_0} + \int^{m_t}_{m_s} \frac{1}{\epsilon_{\rm sc}}\right)dm \right]\frac{dN}{d\log m_t}d\log m_t.
\end{align}
Taking $m_0 = 10^{-4}$ (roughly the mass of Ceres), our $\langle M_{\rm req} \rangle$ are summarized in Table \ref{tab:Mreq}. Increasing $m_0$ by two orders of magnitude makes negligible difference. In general, the required dust mass by our simple model of planetesimal accretion is significantly smaller than that under pebble accretion for FFPs and comparable for the bound planets. The reduction in $\langle M_{\rm req} \rangle$ under planetesimal accretion is expected given that $\epsilon_{\rm eff} < 1$ for {\it higher} planet masses whereas the FFP mass functions (median and `hi') are bottom-heavy. We note that the model for planetesimal accretion we adopt here is necessarily simple. While the true $\epsilon_{\rm sc}$ could be smaller (even more scattering) or higher (source of damping by smaller bodies or small amount of gas; \citealt{Goldreich04}), any $\epsilon_{\rm sc} < 1$ will affect higher mass rather than low mass planets so our statement that $\langle M_{\rm req} \rangle$ under planetesimal accretion is close to $M_{\rm sld}$ is robust to such uncertainties.

Growth by giant impact is more likely to occur in the later stage of assembly process when pebble and/or planetesimal accretion has already contributed to a sizeable growth of the pre-impact bodies. Following the same argument as the previous paragraph, at $\sim$3 au, close encounters between bodies more massive than $\sim$3.65$M_\oplus$ would lead to ejection rather than collisions. Giant impact therefore is more relevant in the question of what shapes the FFP and bound planet mass functions rather than the efficiency of planet formation. In systems of equal mass planets, only the planet pairs with Safronov number $>$1---corresponding to more massive planets---are expected to lead to ejection. It follows that we may expect a flat or top-heavy FFP mass function and a declining bound planet mass function beyond $M_p \gtrsim 3.65 M_\oplus$ \citep{Chachan24}, which is in qualitative agreement with the observation of bound planet mass function but not necessarily with the observed FFPs. A more bottom-heavy FFP mass function can arise in a system of non-equal mass planets whereby the most massive planet scatters and ejects their lighter neighbors, remaining a sole survivor.

Instead of planet-planet scattering, ejection from circumbinary planetary systems through interaction with the central stellar binary has been proposed as a way to reproduce the FFP mass function at high masses \citep{Coleman2025}. Due to the large angular momentum reservoir supplied by tight binary stars, these systems are efficient in scattering a large number of planets onto unbound orbits during planetary migration prior to disk dispersal. However, uncertainty in the effect of binary systems on planet formation rates make a quantitative comparison to necessary initial disk masses challenging in the context of this study.

\begin{figure*}
    \centering
    \gridline{\fig{disk_mf.pdf}{0.45\linewidth}{}
            \fig{disk_mf_planete.pdf}{0.45\linewidth}{}}
    \caption{The cumulative distribution function of disk solid masses from \citet[][dotted, `T Tauri']{Manara23}, \citet[][solid, `Class 0/I']{Tobin20}. Indicated with vertical lines are the total planet masses including both microlensing planets (FFP and BND) and short-period planets with the gray bar spanning the 1-$\sigma$ error on the mass functions. The black horizontal dotted line marks the median CDF = 0.5 while the blue horizontal dotted line represents the fraction of stars with {\it Kepler} planets from \citet{Yang20} for stars of K4 type and later with its 1-$\sigma$ error illustrated with the horizontal gray bar. The colored vertical bars draw the range of {\it required} dust mass accounting for variable planet formation efficiency. Left: pebble accretion---the lower and upper limits of each of the colored bars are set by the seed mass $10^{-2}M_\oplus$ and $10^{-5}M_\oplus$, respectively. The hatched bar uses that of `lo' mass functions. The required dust mass corresponding to `hi' mass functions fall outside the plotting range and are not shown. Right: plantesimal accretion---the lower and upper limits of the colored bar are set by the `lo' and `hi' mass functions, respectively with the vertical dashed colored lines give the median. Top: MOA survey results used for the bound planets. Bottom: KMTnet survey results used for the bound planets with the single power law shown with the dotted vertical line and the blue vertical bar, and the double Gaussian function shown with the dashed vertical line and the magenta vertical bar. The blue and magenta bars overlap with each other.}
    \label{fig:mass-budget}
\end{figure*}

\begin{figure*}
    \centering
    \gridline{\fig{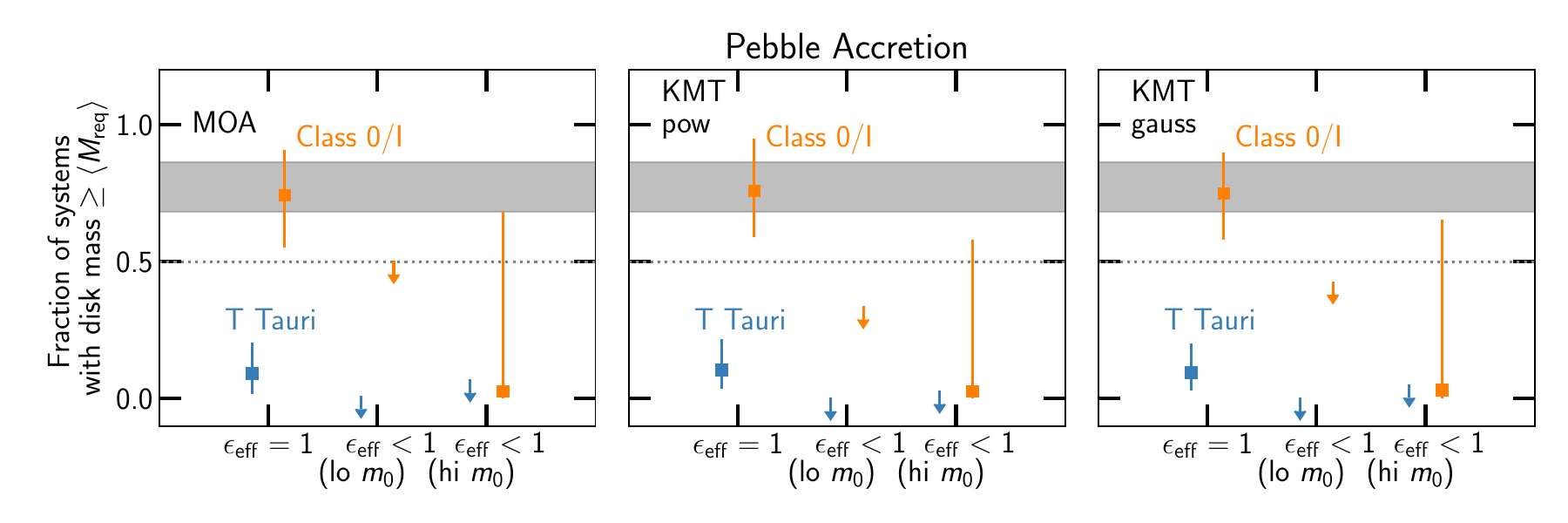}{\linewidth}{}}
    \gridline{\fig{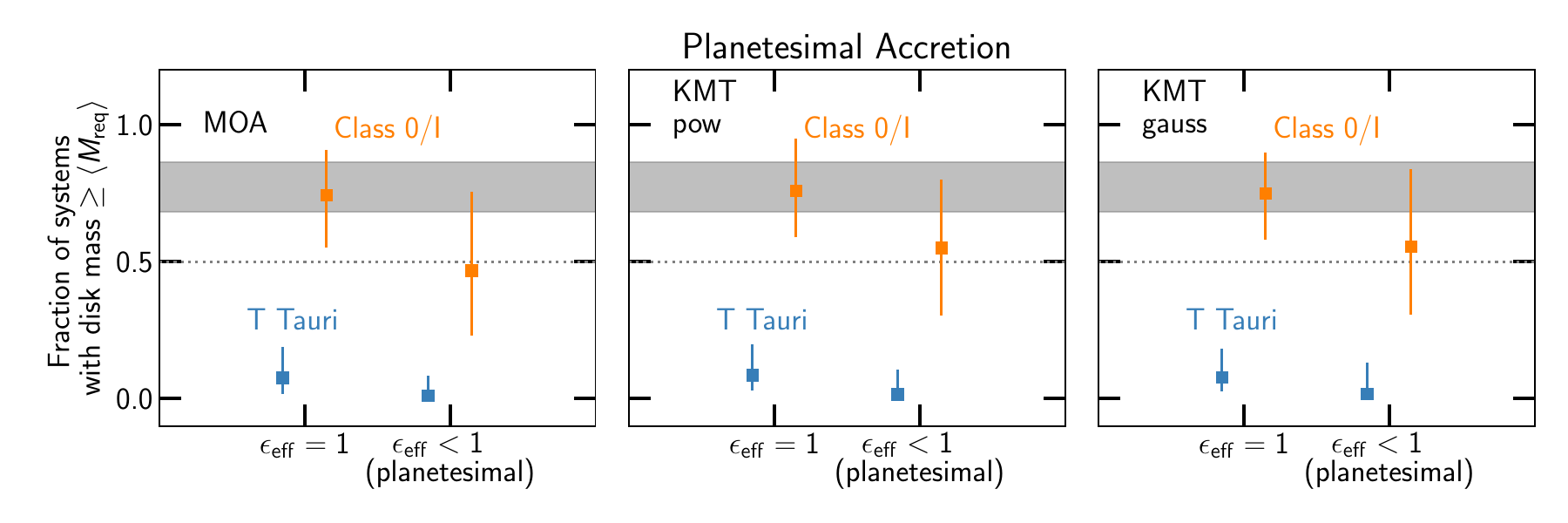}{\linewidth}{}}
    \caption{Fraction of systems with enough mass to produce planets, calculated by reading where the CDFs in Figure \ref{fig:mass-budget} intersect the mean total solid mass per star under $\epsilon_{\rm eff}=1$ and $\epsilon_{\rm eff} < 1$ (the vertical lines and bands in Figure \ref{fig:mass-budget}). From left to right, we show the results for MOA, KMTnet power-law and KMTnet Gaussian planet mass functions for bound planets. In each panel, blue and orange represent underlying disk mass distributions of T Tauri \citep{Manara23} and Class 0/I \citep{Tobin20} respectively. The markers correspond to the median planet mass function while the lower and the upper ends of the errorbar to the `lo' and `hi' planet mass functions, respectively. Only the upper limits are shown if the the required solid mass is larger than the maximum disk mass under the median solid mass function. The gray horizontal bar draws the measured fraction of stars harboring a planet \citep{Yang20} while the horizontal dotted line gives the median 0.5. Top: pebble accretion---`lo $m_0$' and `hi $m_0$' each correspond to seed masses of $10^{-5} M_\oplus$ and $10^{-2} M_\oplus$. Bottom: planetesimal accretion.}
    \label{fig:pfrac}
\end{figure*}

\section{Results}
\label{sec:results}

Table \ref{tab:Mreq} summarizes our $\langle M_{\rm req} \rangle$ for each of the mass functions we explore. Under pebble accretion, starting from smaller seed mass ($m_0$) requires more dust mass, as smaller seeds are less effective in capturing pebbles. The required mass to account for all the FFPs overwhelms that for bound planets by roughly a factor of $\sim$10--20 for the median and `hi' mass functions and a factor of $\sim$2--3 for the `lo' mass function. Although the total $M_{\rm sld}$ is only a factor of $\sim$3 larger for FFPs than the bound planets, FFP median and `hi' mass functions are bottom-heavy, and it is the low planet masses whose $\epsilon_{\rm eff}$ is less than 1. The two facts combine to greatly enhance $\langle M_{\rm req} \rangle$ well beyond the total $M_{\rm sld}$. If the FFP mass function is truly bottom heavy, it implies that on average, planetary systems must have had $\gtrsim$2000 $M_\oplus$ in solids to begin with and the majority of that mass is ``expelled'' by ejecting/launching the FFPs to either unbound or wide-bound orbits.

Such large amounts of required mass may pose a problem. In Figure \ref{fig:mass-budget}, we plot the cumulative distribution function (CDF) of solid masses for T Tauri disks \citep{Manara23} and Class 0/I disks from Orion as reported by \citet{Tobin20}. We overplot the total $M_{\rm sld}$ summing up the FFP, bound, and inner planets assuming 100\% planet formation efficiency with black vertical lines, as well as $\langle M_{\rm req} \rangle$ with shaded colorbars. In the calculation of the latter, we assume all the solids that went into the inner region can effectively be incorporated into the planets so that $M_{\rm sld}$ for the inner planets is equal to their $\langle M_{\rm req} \rangle$. Figure \ref{fig:pfrac} illustrates an alternative view where we report the fraction of systems that have disk mass $\geq$ $M_{\rm sld}$ (100\% efficiency) or $\langle M_{\rm req} \rangle$ for T Tauri and Class 0/I disks.

Even at 100\% planet formation efficiency, we see that only the top $\sim$10--20\% heaviest T Tauri disks can account for the planets so the solids must have originated from an earlier stage where the disk is more massive, reproducing the classic result of \citet{Najita14}. Our result is consistent with that of \citet{Mulders21} in the sense that they could just barely account for all the planets' masses at short periods with T Tauri disks. We are adding the contribution of microlensing planets on top of short-period planets, and with the mass of the former exceeding the mass of the latter, it follows that we run out of mass budget with T Tauri disks. When we consider younger (Class 0/I) disks instead, $\sim$75\% of such disks have enough mass to account for the solid mass required to create a planetary system which not only exceeds the median 50\% but is also high enough to match the observed fraction of stars (of types later than K) harboring a planet computed using {\it Kepler} planets \citep[see][their Figure 5]{Yang20}.

If we account for $\epsilon_{\rm eff} < 1$, then even the heaviest T Tauri disk is insufficient. In fact, under pebble accretion, even Class 0/I disks would not have enough mass if the FFP mass function is bottom-heavy. For at least 50\% of the Class 0/I disks to have enough mass to account for $\langle M_{\rm req} \rangle$, we require the seed mass to be large ($10^{-2}M_\oplus$ rather than $10^{-5}M_\oplus$) and we need the FFP mass function to be peaked (top-heavy on the low mass end) rather than bottom-heavy.

Under planetesimal accretion, we find that T Tauri disks remain too light to account for the planet masses. We find more favorable results with Class 0/I disks where $\gtrsim$50\% of these young disks have enough mass to account for $\langle M_{\rm req} \rangle$ from the median FFP mass function. For more peaked/top-heavy FFP mass function, the fraction of Class 0/I disks with mass $\geq \langle M_{\rm req} \rangle$ matches the fraction of stars with a planet from the {\it Kepler} survey, reconciling disk and exoplanet statistics. We note however that the estimated $\langle M_{\rm req} \rangle$ under planetesimal accretion here may be underestimated as our simple prescription does not take into account the assembly process of the planetesimals themselves, which may be inefficient \citep[e.g.,][]{Carrera22,Li25}. 

In our comparison to exoplanet occurrence rate studies, we use the analysis of the transiting planets from {\it Kepler} because a similar statistics does not exist yet for microlensing planets. Nevertheless, we expect a similar fraction of stars with planets given that there tends to be more mass in the outer disk than in the inner disk, so that if the system was able to create outer planets, it should have no problem creating the inner planets.

\section{Discussion and Conclusion}
\label{sec:discussion}
{he key reason behind the missing mass problem we recover in exoplanet vs.~disk observations is the sheer number of FFPs per star suggested by the bottom-heavy mass function reported in current microlensing observations. This new observation leads to two main findings of our analysis. 

First, the solid mass locked in FFPs is comparable to that in bound and inner planet populations such that their total sum is in excess of the amount of solid material available among T Tauri disks and comparable to the total solid mass reservoir available among Class 0/I disks.

Second, when accounting for the lossy nature of planet-formation via pebble accretion, the solid mass reservoir required to assemble the FFP population greatly enhances and exceeds the mass available around most stars during the Class 0/I phase.

While uncertainties in planet formation can alter the absolute value of planet formation efficiency, gravity dictates that planetary growth will be more sluggish when the planet is less massive, so as long as the FFP mass function is bottom-heavy, the missing mass problem will likely persist. We close this paper with the discussion of the implications of our results and motivation for future investigation.

\subsection{Class 0/I Disks as Sites of Planet Formation}
\label{ssec:timing}

With the inclusion of microlensing planets, we find that T Tauri disks have insufficient mass to account for the typical solid mass in planets per star, even at 100\% planet formation efficiency, suggesting that the {\it initial} build-up of planets begins from the early stages of protoplanetary disks.\footnote{The nature of the {\it onset} of planet formation can be in the form of early planetesimal formation leading to early core coagulation that triggers early pebble accretion. Our analysis is agnostic to the exact process as we simply count the total dust mass in young disks that gives the {\it initial} solid mass budget.} However, when the variable planet formation efficiency under pebble accretion is taken into account, we find that even Class 0/I disks are too light by at least an order of magnitude.\footnote{While it has been argued that the published masses of T Tauri disks are underestimated due to uncertain optical properties \citep[e.g.,][]{Savvidou25,Savvidou26}, the suggested corrective factors of $\sim$2--7 from the formation of dust substructures \citep[e.g.,][]{Godines26} will not be enough, as we demonstrate that even much more massive Class 0/I disks are insufficient to explain the required solid masses to account for all the planets observed.} Accounting for planetaesimal accretion alleviates the problem of mass budget, with a caveat that if the fraction of stars that harbor a planet remains just as high for microlensing planets as {\it Kepler} planets, then either the planet mass function cannot be bottom-heavy or Class 0/I disks remain deficient in mass.

Class 0/I disks are embedded in protostellar cores so we may expect continuous accretion of dust material from the surrounding medium onto the disk. It is difficult however to consider this external accretion as a solution to the systematic problem of the missing mass budget. For example, \citet{Sheehan22} run radiative transfer models for select Class 0/I disks in the sample from \citet{Tobin20}, self-consistently solving for the disk and the envelope component. Their quoted envelope-to-disk dust mass ratio range from $\sim$10$^{-2}$ to $\sim$3100 with a median value of $\sim$0.62, suggesting that in at least half the cases, the dust mass in the envelope is comparable or smaller than that in the disk. 

It may be that these young disks are in fact massive enough to be marginally gravitationally stable. 
In converting the dust thermal emission into masses, several assumptions are made including the source of heat, whether the dust is optically thick or thin, and the temperature structure of the disk. \citet{Xu22} present revised mass conversion using the same dataset of \citet{Tobin20} relaxing the assumption of vertical isothermal disk, optically thin grains and allowing for internal accretional heating and marginal gravitational instability. Fitting over two wavelength measurements, \citet{Xu22} find significantly higher dust masses (see their Figure 15), roughly $\sim$7 times higher masses compared to \citet{Tobin20}. Such heavy disks are, by construction, uncomfortably on the edge of gravitational instability (Toomre Q $\sim$1--2).

When we boost the disk masses reported in \citet{Tobin20} by a factor of 7 to approximate the result of \citet{Xu22}, we are able to reconcile the required dust masses to generate the observed planets per system with Class 0/I disks, although the FFP mass function would still need to be less bottom-heavy than the median mass function we use. The physical assumption behind the result of \citet{Xu22} is that the disk self-regulates to be at a state of marginal gravitational stability as the copious amount of infall from the envelope is balanced by the gravitational transport of angular momentum \citep[e.g.,][]{Vorobyov07}. In this sense, the condition of marginal stability may be an expected outcome should most of the mass be locked into the protostellar core and need to be transported down to the circumstellar disk during the Class 0/I phase.

\subsection{Different Formation Environments for Free-Floating versus Bound Planets}

Another way to approach the problem of missing mass budget is to consider different formation environment of free-floating and bound planets. If indeed the FFP mass function is more bottom-heavy, then the total required mass to create FFPs exceeds that of bound planets by at least an order of magnitude under pebble accretion and by factors of $\sim$2--4 under planetesimal accretion (see Table \ref{tab:Mreq}). Considering just the bound planets would bring the fraction of disks with enough mass to create planets into better agreement with the fraction of stars (later types than K) with {\it Kepler} planets. The same occurrence rate studies report the fraction of stars with planets drops sharply to $\sim$30\% for Sun-like and hotter stars \citep{Yang20}. 

Given how the mass required to create FFPs per star can only be reconciled with the most massive disks and more massive stars tend to have more massive disks \citep{Manara23}, it may be that the FFPs are preferentially created around more massive stars consuming the majority of the mass budget there, leading to a lower fraction of such stars with currently observable bound planets, in qualitative agreement with {\it Kepler}-based statistics. On the other hand, if the FFP mass function is more peaked or top-heavy on the low mass end, FFPs and bound planets may form in similar environments with no preference for stellar mass.

\subsection{Solid Masses Not Accounted For}

Our estimation of the total mass of solids incorporated into planetary objects per system is necessarily limited by the current observational sensitivities. 
For example, in the inner regions of planetary systems, RV and transit surveys are generally insensitive to planets with mass or radius less than Earth. By using the minimum mass extrasolar nebula constructed by \citet{Dai20} in computing the inner planet total mass budget, we inherit their uncertainties: namely, the lack of bias correction on the RV mass measurements (as the dataset is inhomogeneous) and the exclusion of planets of size $<$1R$_\oplus$ from the transit survey. A substantial reservoir of lower-mass planets and leftover planetesimals could exist.   

Similarly, current microlensing surveys have limited sensitivity to planets with masses below that of Earth \citep[e.g.,][]{Zang25}, leaving open the possibility of a significant reservoir of cold, bound, very low-mass planetary objects. Our analysis is an extrapolation of the reported microlensing mass functions down to 0.33$M_\oplus$, but the cold planet mass function could extend to even lower masses (and the shape of the function could be different), as we discuss in more detail in Section \ref{ssec:space-microlens}. 
If such very low mass objects are just as numerous as the population of FFPs with $M_p\gtrsim 0.33M_\oplus$ inferred from current microlensing surveys that we adopt here, the missing mass problem explored here will become even more dire.

There may also be a substantial population of low-mass objects bound to stars at large separations, analogous to the inner and outer Oort clouds in our solar system. Objects with semimajor axes greater than tens of au would appear as free-floating planet (FFP) events in microlensing surveys.  
Even in our own solar system, the total mass of the inner and outer Oort clouds is highly uncertain, with estimates ranging from $\sim 1-100~M_\oplus$, and a most likely value of a few to $\sim$10$M_\oplus$ \citep{Menichella2026}. Owing to its well-characterized selection function, the Vera C.~Rubin Observatory (Rubin) is expected to improve constraints on the number of Oort Cloud objects \citep[e.g.,][]{Solontoi2010,Inno2024}. If typical exoplanetary systems also have an Oort cloud analogue of similar masses, their mass contribution would be at most comparable to the FFP and bound planet mass from microlensing surveys estimated in this paper assuming $\epsilon_{\rm eff} = 1$. It remains unclear, however, whether the solar system is representative of planetary systems more generally.

For FFPs, we have limited our consideration to that from microlensing surveys, which are currently sensitive only to objects with masses $\gtrsim$1$M_\oplus$. Another potential contributor is a population of interstellar interlopers, for which there are three detections so far: 1I/`Oumuamua \citep{Meech2017}, 2I/Borisov \citep{Guzik2020,Jewitt2019}, and 3I/Atlas \citep{Bolin2025}. \citet{Do2018} estimate the sensitivity of the Pan-STARSS survey to objects like ‘Oumuamua, inferring a number density of such objects of $\sim$10$^{15}$--10$^{16}$ pc$^{-3}$ per star given the number density of stars of $0.1~{\rm pc^{-3}}$ in the local solar neighborhood. The mass of `Oumuamua is very uncertain, but assuming a mass density of $3~{\rm g~cm^{-3}}$ and dimensions of 200~m $\times$ 20~m $\times$ 20~m \citep{Fitzsimmons2018}, we estimate a mass of $2.4 \times 10^{11}~{\rm g} \simeq 4\times 10^{-17}~M_\oplus$ and thus a total mass of $\sim$0.4$M_\oplus$ per star, which would be a negligible contribution to the total solid mass budget. Borisov and Atlas are likely more massive, but they were discovered in surveys with poorly characterized selection functions, making it difficult to use them to estimate the mass density of similar objects.

If the interstellar interlopers were of the same population as the FFPs measured by microlensing surveys, the total mass budget would be more substantial. For instance, \citet{Gould2022} estimate 0.75$M_\oplus$ per decade of mass per solar mass of stars worth of `Oumumua-like objects of mass 3$\times$10$^{-17}M_\oplus$. Evaluating our median FFP mass function (equation \ref{eq:dNdMp_ffp}) at $M_p = 3\times 10^{-17}M_\oplus$ and multiplying by the same $M_p$, we obtain $\sim$3.5$M_\oplus$ per decade of mass per star worth of `Oumuamua-like objects. Based on the similar numbers, we could conjecture that the FFP mass function is continuous down to these extremely low masses \citep[see, e.g.,][]{Gould2022} which would imply a total solid mass of $\sim$170$M_\oplus$ of FFPs down to $M_p = 3\times 10^{-17}M_\oplus$ at $\epsilon_{\rm eff}=1$. This substantial addition of extra mass budget would severely exacerbate the missing mass problem; however, we consider a continuous distribution of masses over 16 orders of magnitude unlikely. Simulations of streaming instability report the initial mass function of planetesimals likely follows a broken power law with a peak at $\sim$6$\times 10^{-8} M_\oplus$ \citep[see][their Run I, the location of the break is at a similar mass in their Run II]{Li19} such that there is more mass in larger planetesimals (see their Figure 4). Furthermore, some of these planetesimals are likely incorporated into the final planets. Improved constraints on the number density of interstellar interlopers by Rubin would nevertheless be useful \citep{Hoover2022}.

Our analysis also does not explicitly consider the directly imaged planets. Their masses and orbital distances have some overlap with the current microlensing surveys but approximately half of them also fall outside of the microlensing sensitivity limit \citep{Zhu21}. It is not trivial to properly account for this overlap in the overall mass budget and \citet{Yee25} report the occurrence rate of directly imaged objects in the mass range $\sim$1--80$M_{\rm Jup}$ \citep{Nielsen19,Vigan21} lie above that expected from microlensing FFPs (see their Figure 4) suggesting many of the directly imaged objects are part of the extension of the stellar and brown dwarf population. More importantly, while typical microlensing planets are thought to be around stars of mass $\sim$0.56$M_\odot$, directly imaged bound planets are preferentially found around high mass stars $>$1.5$M_\odot$ \citep[see, e.g.,][their Figure 6]{Nielsen19}. Even though these massive planets are also observed to be metal-enriched (e.g., HR 8799 planets are inferred to have collectively $\sim$400$M_\oplus$ worth of solids; \citealt{Ruffio26,Xuan26}), given their low integrated occurrence rate $\sim$9\% around high mass stars, which themselves are rare \citep{Nielsen19}, we consider their overall budget to the {\it typical} stars given our stellar initial mass function to be minor. 

Another point to consider is that the aforementioned direct imaging surveys are that of bound objects, not the free-floaters. While there have been measurements of free-floating objects through direct imaging \citep[e.g.,][]{Bouy22,Miret-Roig22}, their planetary vs.~stellar nature remains elusive. Nevertheless, the problem of mass budget we identify in this paper stems from the oversized contribution of the low mass ($\lesssim M_\oplus$) planets so our results are robust to the uncertainties of the mass functions at the high mass end at a few Jupiter masses where the distinction between planets and substellar objects blurs.

There are non-planetary contributions such as moons, comets, and debris belts which contribute negligibly in our solar system ($\lesssim$0.1$M_\oplus$) \citep{Menichella2026}. Planets may also be accreted onto the host star after planet formation is largely complete, by e.g., planet-planet scattering. Based on the observed abundance anomalies in the photospheres of stars, \citet{Liu2024} estimate an average of $\sim$5$M_\oplus$ of rocky material is accreted by the $\sim 8\%$ of stars that show evidence for this accretion, or $\sim$0.5$M_\oplus$ per star averaged over all stars, which is a negligible contribution. Furthermore, even if a pair of stars is considered to be co-moving or have originated from the same star-forming region, spatial variations in metallicity within the molecular cloud can also give rise to the different photospheric abundances \citep{Soliman2025}.

\subsection{The Case for Space-Based Microlensing Surveys}
\label{ssec:space-microlens}

It is likely that the planet formation efficiency is not unity irrespective of the assembly process. In such a scenario, a bottom-heavy planet mass function (especially the FFP mass function with its higher normalization than the reported bound mass functions) is problematic as the required dust mass to account for all the planets exceeds the amount of mass available in protoplanetary disks, even in the Class 0/I phase.

At the time of writing this paper, the shape of the mass function for both cold bound and FFP planets from microlensing is poorly constrained.  For bound planets, the study by \citet{Zang25} only included 6 planets with mass ratio $<$10$^{-4.5}$, corresponding to $M_p \lesssim 5.9~M_\oplus$ for the average stellar mass of $0.56~M_\odot$, and the lowest mass ratio was $q=10^{-5.2}$ corresponding to a mass of $\sim 1.2~M_\oplus$. The earlier study by \cite{Suzuki16} did not include any events with mass ratio below $10^{-4.5}$, with the lowest mass ratio being $10^{-4.2}$.  

For FFPs, the mass function derived by \citet{Sumi23} was based on the sample of \citet{Koshimoto2023}, which only had 12 events with timescales that were $<$1 day, corresponding to low-mass lenses. Of these, only two exhibit significant finite source effects that allow for a more robust estimation of the lens mass through a measurement of the angular Einstein radius $\theta_{\rm E}$: MOA-9y-5919 with $\theta_{\rm E}=0.9\pm 0.4~{\rm \mu}{\rm as}$ corresponding to $0.75^{+1.23}_{-0.46}M_\oplus$ \citep{Koshimoto2023} and OGLE-2016-BLG-1928 with $\theta_{\rm E} = 0.842 \pm 0.064~\mu$as corresponding to $\sim$0.3--2$M_\oplus$ \citep{Mroz2020}. The FFP mass function derived by \citet{Gould2022} was essentially based on only four events with measurements of $\theta_{\rm E} \sim$4--10$\mu$as, and therefore likely not terrestrial.

Any statements about the mass function for planets with mass significantly less than $\sim$1$M_\oplus$ are therefore based solely on extrapolation.  
Limiting the minimum $M_{\rm sld}$ to 1$M_\oplus$ lowers $\langle M_{\rm req} \rangle$ for the FFPs under {\it bottom-heavy} mass function by factors of $\sim$3 for pebble accretion (for all other cases such as top-heavy mass function or for planetesimal accretion, the reduction in $\langle M_{\rm req} \rangle$ is less than factors of 2).\footnote{The minor numerical correction is not surprising given that 1$M_\oplus$ is not that different from our adopted lower limit 0.33$M_\oplus$.} This level of reduction is inadequate to bring the fraction of Class 0/I disks with enough mass to account for $\langle M_{\rm req} \rangle$ up to $\sim$50--75\% so the problem of mass budget remains {\it if} the FFP mass function is bottom-heavy. 

Future space-based microlensing surveys will provide much stronger constraints on the mass function of cold bound planets and FFPs. In particular, the Nancy Grace Roman Space Telescope (Roman) and the Earth 2.0 (ET) telescope will be instrumental in constraining the bottom of the planet mass function.

As one of the Core Community Surveys to be carried out by Roman \citep{Spergel2015}, the Roman Galactic Bulge Time Domain Survey (RGBTDS) will monitor approximately 1.4 square degrees near the Galactic center using five Roman Wide Field Instrument (WFI) fields, with a cadence of $\sim$12 minutes over a span of six seasons of 72 days each, for a total duration of 432 days for the high-cadence seasons. As one science focus of the RGBTDS, the Roman Galactic Exoplanet Survey (RGES) will detect $\sim 50{,}000$ microlensing events, which will include thousands of cold, bound planets with masses down to that of Ganymede \citep{Penny19}. 
Extrapolating the power-law mass function of \citet{Zang25} (equation \ref{eq:dNdq_zang_pl}) down to the mass of the moon (0.01$M_\oplus$, or equivalently, $\log q = -7.3$), we estimate based on the yields of \citet{Penny19} that RGES should detect $\sim$30 Mars-mass planets and $\sim 10$ lunar-mass objects.\footnote{Note that our estimate of the yield of lunar-mass objects is based on extrapolating Table 3 of \citet{Penny19}. Our estimated yield is likely overestimated because finite source effects preferentially suppress the amplitudes of planetary deviations at such low masses.}
If instead the mass function plateaus at $\sim 1$ planet per dex per star below $M_\oplus$, RGES will only detect $\sim 10$ Mars-mass planets and $\sim 1$ lunar-mass object. 

The RGES will also have sensitivity to widely-bound planets and FFPs with masses of Mars or even below \citep{Johnson2020}. Assuming the power-law FFP mass function of \citet{Sumi23} extends down to $\sim$0.01$M_\oplus$, RGES will detect $\sim$400 FFPs with $M_p\sim M_\oplus$, $\sim$750 with $M_p\sim 0.1 M_\oplus$, and $\sim$470 with $M_p\sim 0.01 M_\oplus$. The \citet{Sumi23} FFP mass function (equation \ref{eq:dNdMp_ffp}) implies roughly equal total mass in FFP per dex in mass of $\sim$10$M_\oplus$. From the inspection of Figure 8 of \citet{Johnson2020}, if equation \ref{eq:dNdMp_ffp} holds, RGES will have sensitivity to FFP with masses down to $\sim$0.002$M_\oplus$, or roughly the mass of Pluto. With a significantly larger number of detections, Roman will enable a strong constraint on the mass function of bound and free-floating planets down to sub-Earth masses.

Earth 2.0 (ET) is an approved space mission that will use seven 30 cm telescopes to survey for exoplanets using transits and microlensing \citep{Ge2022}.\footnote{https://et.shao.ac.cn/}  One of these telescopes will be used to monitor a 4 deg$^2$ field toward the Galactic center for a total of 2 years.  The same field will also be monitored simultaneously by the ground-based KMTNet survey \citep{Kim2016}. \citet{Ge2022} predict that ET+KMTNet will detect $\sim$600 FFP events, assuming a power-law distribution for the FFP population with a somewhat lower normalization but a slope similar to \cite{Sumi23} for $M_p \geq M_\oplus$, but a constant frequency per dex in mass below this mass. Assuming their FFP distribution and extending it to $0.1~M_\oplus$, \citet{Sumi23} predict that ET+KMTNet should detect $\sim 840$ FFP planet events, including $\sim 210$ with mass $\le M_\oplus$. Thus, ET+KMTNet will also provide strong constraints on the low-mass end of the FFP mass function.  

Perhaps more importantly, terrestrial FFP events that are simultaneously detected by ET and KMTNet will have measurably different peak times and impact parameters due to the fact that the baseline between the KMTNet's location on Earth and ET's location at L2 of $\sim 0.01$~au is roughly commensurate with the size of the Einstein ring of a $M_\oplus$ lens projected onto the observer plane \citep{Gould03}. This setup allows for the measurement of the microlensing parallax $\pi_{\rm E}$, which, when combined with $\theta_{\rm E}$ from the morphology of the FFP light curves due to the finite size of the source, allows for a direct geometric determination of the mass of the planet \citep{Dong2026}. Assuming the \citet{Sumi23} mass function and using the yields from \citet{Ge2022}, we estimate that ET+KMTNet will measure the masses of $\sim 300$ FFP planets, including $\sim 175$ with $0.1~M_\oplus \le M_p \le M_\oplus$, therefore enabling a robust, direct measure of the frequency of low-mass FFPs.\footnote{\citet{Ge2022} did not estimate the yield of mass measurements with ET+KMTNet for FFPs with mass below $0.1~M_\oplus$. However, as these mass measurements rely on detecting and characterizing the events from both space (with ET) and the ground (with KMTNet), and the sensitivity of KMTNet falls off quickly for planets with mass below $\la$1$M_\oplus$ (see Fig.~S1 of \citealt{Zang25}), it is unlikely that the combined ET+KMTNet survey will measure the masses of a large number of planets with mass substantially less than 0.1$M_\oplus$.}

The space-based microlensing results may prove that there is no missing mass problem in planet formation---or they may reaffirm the problem is real and in need of resolution.

\begin{acknowledgments}
We thank Naoki Koshimoto and Daisuke Suzuki for their help in understanding microlensing mass function statistics. Ilse Cleeves and John Tobin provided useful comments on the disk mass measurements. We further acknowledge useful discussions with Darryl Seligman and Michael Tucker on the mass budget of interstellar interlopers, and Nadine Soliman on the contribution of accreted solid material onto the star to the mass budget.
This project was conceived at the Rogue Worlds meeting in Osaka, Japan, in December 2024. We thank the organizers of that meeting. EJL was supported by NSF Research Grant 2509275, NSERC Discovery Grant RGPIN-2020-07045, DGECR-2020-00230, and the William Dawson Scholarship from McGill University. WD was supported by NSF grant PHY-2210361 and the Maryland Center for Fundamental Physics. 
S.H. acknowledges support by the Natural Sciences and Engineering Research Council of Canada (NSERC), funding references CITA 490888-16 and RGPIN-2020-03885.
BSG was supported by National Aeronautics and Space Administration grant 80NSSC24M0022 and by The Ohio State University through the Thomas Jefferson Chair for Discovery and Space Exploration endowment.
\end{acknowledgments}

\begin{contribution}
EJL led the calculations and the writing of the manuscript. WD recalculated the confidence intervals of microlensing mass functions, checked the initial calculations, and edited the manuscript. SH checked the initial calculations and edited the manuscript. BSG asked the original question of the mass budget in light of microlensing mass functions, contributed to the text, and edited the manuscript.
\end{contribution}

\bibliography{ffp}{}
\bibliographystyle{aasjournalv7}
\end{document}